\documentclass[prd,twocolumn,showpacs,amsmath,amssymb,floatfix]{revtex4}
\usepackage{graphicx,color,dcolumn,booktabs}
\usepackage{longtable,lscape}
\usepackage{txfonts}
\usepackage{amssymb}
\usepackage{indentfirst}

\begin{document}
\title{Novel charmonium-like structures in the $J/\psi\phi$ and $J/\psi\omega$ invariant mass spectra}

\author{Xiang Liu$^{1,2}$
\footnote{Corresponding authors.} } \email{xiangliu@lzu.edu.cn}

\affiliation{$^1$School of Physical Science and Technology, Lanzhou University, Lanzhou 730000,  China\\
$^2$Research Center for Hadron and CSR Physics, Lanzhou University
$\&$ Institute of Modern Physics of CAS, Lanzhou 730000, China}
\author{Zhi-Gang Luo}
\author{Shi-Lin Zhu\footnote{Corresponding authors.}}\email{zhusl@pku.edu.cn}
\affiliation{Department of Physics
and State Key Laboratory of Nuclear Physics and Technology\\
Peking University, Beijing 100871, China}

\date{\today}
\begin{abstract}

Stimulated by the new evidence of $Y(4274)$ observed in the
$J/\psi\phi$ invariant mass spectrum, we first propose the
charmonium-like state $Y(4274)$ as the S-wave
$D_s\bar{D}_{s0}(2317)+h.c.$ molecular state with $J^P=0^-$, which
is supported well by dynamics study of the system composed of the
pseudoscalar and scalar charmed mesons. The S-wave
$D\bar{D}_{0}(2400)+h.c.$ molecular charmonium appears as the
molecular partner of $Y(4274)$, which is in accord with the
enhancement structure appearing at 4.2 GeV in the $J/\psi\omega$
invariant mass spectrum from $B$ decays. Our study shows that the
enhancement structures, $i.e.$, the newly observed $Y(4274)$ and
the previously announced $Y(4140)/Y(3930)$ in the $J/\psi\phi$ and
$J/\psi\omega$ invariant mass spectra, can be understood well
under the uniform framework of the molecular charmonium, which can
be tested by future experiments.

\end{abstract}

 \maketitle

Very recently the CDF Collaboration \cite{collaboration:2010aa}
studied the $J/\psi \phi$ invariant mass spectrum in the $B\to
J/\psi\phi K$ channel based on the sample of $p\bar{p}$ collision
data with an integrated luminosity of 6 fb$^{-1}$. Besides
confirming the previous $Y(4140)$ state \cite{Aaltonen:2009tz},
CDF also reported the observation of an explicit enhancement
structure with $3.1\sigma$ significance in the $J/\psi \phi$
invariant mass spectrum, which is of mass
$M=4274.4^{\pm8.4}_{-6.7}(\mathrm{stat})$ MeV and width
$\Gamma=32.3^{+21.9}_{-15.3}(\mathrm{stat})$ MeV
\cite{collaboration:2010aa}. We will refer to this new structure
by the name $Y(4274)$ in this letter.

The appearance of $Y(4274)$ in the $J/\psi \phi$ invariant mass
spectrum not only makes the charmonium-like family abundant, but
also raises our interest in exploring the origin of enhancement
structures in the $J/\psi \phi$ invariant mass spectrum and
revealing the relation between $Y(4274)$ and $Y(4140)$, which will
be helpful to improve our knowledge of the underlying properties
of charmonium-like state.

\begin{center}
\begin{figure}[htb]
\begin{tabular}{c}
\scalebox{0.47}{\includegraphics{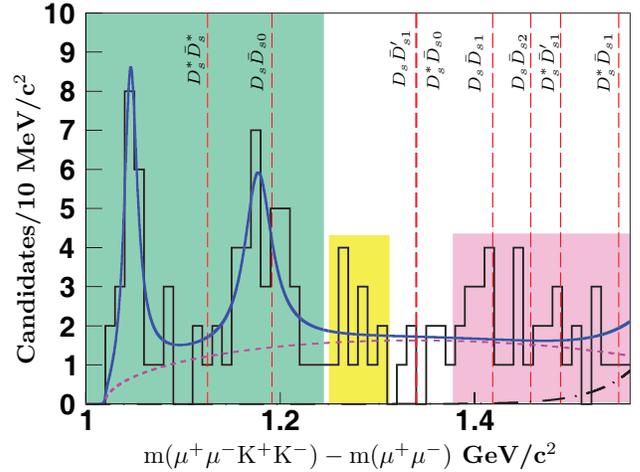}}
\end{tabular}
\caption{(Color online.) The mass difference $\Delta
M=m(\mu^+\mu^-K^+K^-)-m(\mu^+\mu^-)$ distribution   (histogram)
for events in the $B^+$ mass window \cite{collaboration:2010aa}.
Besides $Y(4140)$, one explicit enhancement appears around $4274$
MeV. Here, the purple dashed line is the background from the
three-body phase space. The blue solid line is the fitting result
with resonance parameters of $Y(4140)$ and $Y(4270)$ resonances in
Ref. \cite{collaboration:2010aa}. The vertical red dashed lines
denote the thresholds of $D_s^*\bar{D}_s^*$,
$D_{s}\bar{D}_{s0}(2317)$, $D_{s}\bar{D}_{s1}^\prime(2460)$,
$D_{s}^*\bar{D}_{s0}(2317)$, $D_{s}\bar{D}_{s1}(2536)$,
$D_{s}\bar{D}_{s2}(2573)$, $D_{s}^*\bar{D}_{s1}^\prime(2460)$ and
$D_{s}^*\bar{D}_{s1}(2536)$.  \label{BW}}
\end{figure}
\end{center}

The previous observation of $Y(4140)$ has stimulated great
interest among theorists, especially when associating it with
$Y(3930)$ reported by the Belle Collaboration \cite{Abe:2004zs}
and confirmed by the BaBar Collaboration \cite{Aubert:2007vj}.
Both $Y(4140)$ and $Y(3930)$ were observed in the mass spectrum of
$J/\psi+{light\, vector\, meson}$ in $B$ meson decay
\begin{eqnarray*}
B\to K+\Bigg\{\begin{array}{cc} \underline{J/\psi \phi} & \Longrightarrow Y(4140)\\
\underline{J/\psi \omega} &\Longrightarrow
Y(3930)\end{array}.\label{channel}
\end{eqnarray*}
Generally in the weak decays of $B$ meson, the $c\bar{c}$ pair
creation mainly results from the color-octet mechanism.
Furthermore, a color-octet $q\bar{q}$ pair is easily popped out by
a gluon. Thus, $c$ and $\bar{c}$ capture $\bar{q}$ and $q$
respectively to form a pair of charmed mesons. By this mechanism,
a pair of the charm-strange mesons with the low momentum easily
interact with each other and even form the molecular charmonium.
Additionally, $Y(4140)$ and $Y(3930)$ are close to the thresholds
of $D_s^*\bar{D}_s^*$ and $D^*\bar{D}^*$ respectively, and satisfy
an almost exact mass relation
\begin{eqnarray}
M_{Y(4140)}-2M_{D_s^*}\approx
M_{Y(3930)}-2M_{D^*}.\label{relation}
\end{eqnarray}
The mass difference between $Y(4140)$ and $Y(3930)$ is
approximately equal to that between $\phi$ and $\omega$ mesons:
$M_{Y(4140)}-M_{Y(3930)}\sim M_\phi-M_\omega.$ The peculiarity of
$B\to K (c\bar{c})$ and the similarity between $Y(4140)$ and
$Y(3930)$ provoke an uniform molecular charmonium picture to
reveal the underlying structure of $Y(4140)$ and $Y(3930)$
\cite{Liu:2009ei,Liu:2008tn}. Applying $D_s^*\bar{D}_s^*$ and
$D^*\bar{D}^*$ molecular structures to explain $Y(4140)$ and
$Y(3930)$ respectively not only solves a long-standing puzzle of
the structure of $Y(3930)$, but also opens a window to investigate
the hadron dynamics of exotic state beyond the conventional
$q\bar{q}$ and $qqq$ states. A series of research work related
with $Y(4140)$ were carried out later
\cite{Liu:2009ei,Liu:2008tn,Mahajan:2009pj,Wang:2009ue,Branz:2009yt,
Albuquerque:2009ak,Liu:2009iw,Ding:2009vd,Zhang:2009st,vanBeveren:2009dc,
Stancu:2009ka,Liu:2009pu,Wang:2009ry,Drenska:2009cd,Molina:2009ct}.

In Fig. \ref{BW}, we present the comparison between the
experimental data \cite{collaboration:2010aa} and the thresholds
of the charmed-strange meson pairs. $Y(4274)$ is just below the
threshold of $D_s\bar{D}_{s0}(2317)$ similar to the situation of
$Y(4140)$, which stimulates us to deduce naturally that $Y(4274)$
enhancement results from an S-wave $D_{s}\bar{D}_{s0}(2317)+h.c.$
molecular system $Y^{s\bar{s}}$ with the flavor wave function
\begin{eqnarray}
|Y^{s\bar{s}}\rangle
&=&\frac{1}{\sqrt{2}}\Big[|D_s^+D_{s0}^-\rangle + |D_s^-
D_{s0}^+\rangle\Big].
\end{eqnarray}
The $C$ parity of the isoscalar $Y(4274)$ is positive due to the
$Y(4274)\to J/\psi\phi$ decay mode observed by CDF. As the cousin
of $Y^{s\bar{s}}$, $Y^{u\bar{u}/d\bar{d}}$ is of the flavor wave function
\begin{eqnarray}
|Y^{u\bar{u}/d\bar{d}}\rangle &=&\frac{1}{2}\Big[|\bar{D}_0^0
D^0\rangle+|D^0_0
\bar{D}^0\rangle+|D_0^-D^+\rangle+|D_0^+D^-\rangle\Big].
\end{eqnarray}
For such S-wave pseudoscalar-scalar systems, their quantum number
must be $J^P=0^{-}$. Performing dynamical investigations of
$Y^{s\bar{s}}$ and $Y^{u\bar{u}/d\bar{d}}$ can answer whether
there exist $Y^{s\bar{s}}$ and $Y^{u\bar{u}/d\bar{d}}$ molecular
systems, which is one of the main tasks of this letter. What is
more important is that understanding the underlying structure of
$Y(4274)$ will be helpful for revealing the properties of $Y(4140)$
\cite{Liu:2009ei,Liu:2008tn} taking into account the similarities
between $Y(4274)$ and $Y(4140)$.

Using the effective Lagrangian in the heavy meson chiral
perturbation theory (HM$\chi$PT)
\cite{Yan:1992gz,Casalbuoni:1996pg} and the method developed in
literature \cite{Liu:2007bf}, we obtain the effective potentials
of $Y^{s\bar{s}}$ and $Y^{u\bar{u}/d\bar{d}}$ states
\cite{Shen:2010ky}
\begin{eqnarray}
\mathfrak{V}_{eff}^{s\bar{s}}(r)&=&
V_{\phi}^{Direct}(r)+\frac{2}{3}V_{\eta}^{Cross}(r),\label{ss}\\
\mathfrak{V}_{eff}^{u\bar{u}/d\bar{d}}(r)&=&\frac{3}{2}V_{\rho}^{Direct}(r)+\frac{1}{2}V_{\omega}^{Direct}(r)+
V_{\sigma}^{Direct}(r)\nonumber\\&&+\frac{3}{2}V_{\pi}^{Cross}(r)
+\frac{1}{6}V_{\eta}^{Cross}(r)\label{qq}.
\end{eqnarray}
Here, the subscript of the sub-potential denotes the exchanged
light meson. The general expressions of the sub-potentials
corresponding to the pseudoscalar, sigma and vector meson
exchanges are
\begin{eqnarray}
V^{Direct}_{V}(r) &=& -\frac{\beta\beta^\prime}{2} g_V^2 Y(\Lambda,q_0=0,m_{V},r),\\
V^{Direct}_\sigma (r)&=& -g_\sigma g_{\sigma}^\prime
Y(\Lambda,q_0=0,m_\sigma,r),\\
V^{Cross}_{P}(r) &=& \frac{h^2{q_0^\prime}^2}{f_\pi^2}
Y(\Lambda,q_0^\prime,m_{P},r),
\end{eqnarray}
where $f_\pi=132$ MeV and $g_V=m_\rho/f_\pi=5.8$. $g_V$, $h$,
$\beta^{(\prime)}$, $g_\sigma^{(\prime)}$ are the parameters in
the effective Lagrangian, which describe the interaction of the
heavy flavor mesons with the light mesons
\cite{Casalbuoni:1996pg}. $q_0^{\prime}$ is taken as
$m_{D_{s0}}-m_{D_s}$ and $m_{D_{0}}-m_{D}$ for $Y^{s\bar{s}}$ and
$Y^{u\bar{u}/d\bar{d}}$, respectively. And the $Y$ function is
\begin{eqnarray*}
&&Y(\Lambda, \kappa, m, r)  \nonumber\\
&&= \left\{ \begin{array}{ll}
            \mbox{if $|\kappa|\le m$},\quad -\frac{1}{4\pi
              r}\left( e^{-\zeta_1r} -
            e^{-\zeta_2r}\right) +
            \frac{1}{8\pi}\frac{\zeta_2^2-\zeta_1^2}{\zeta_2}
            e^{-\zeta_2r} &\\
            \mbox{otherwise},\quad -\frac{1}{4\pi r}\left(
            \cos(\zeta_1^\prime r)- e^{-\zeta_2r}\right) +
            \frac{1}{8\pi}\frac{\zeta_2^2+\zeta_1^{\prime 2}}{\zeta_2} e^{-\zeta_2r} &
            \end{array}
     \right.
\end{eqnarray*}
with $\zeta_1=\sqrt{m^2-\kappa^2}$,
$\zeta_1^\prime=\sqrt{\kappa^2-m^2}$ and
$\zeta_2=\sqrt{\Lambda^2-\kappa^2}$. $\Lambda$ is the cutoff to
cure the singularity of the effective potential.

\begin{center}
\begin{figure}[htb]
\begin{tabular}{cc}
\scalebox{0.9}{\includegraphics{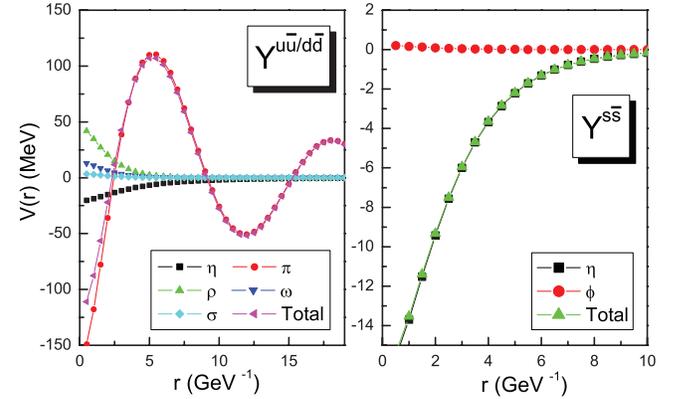}}
\end{tabular}
\caption{(Color online.) The potentials of $Y^{s\bar{s}}$
(right-side diagram) and $Y^{u\bar{u}/d\bar{d}}$ (left-side
diagram) with typical value $\Lambda=1$ GeV. Here, we take
$\beta=0.9$, $\beta^\prime=1$, $g_\sigma=g_\sigma^\prime=-0.76$,
$h=-0.56\pm 0.28$ following Refs.
\cite{Casalbuoni:1996pg,Bardeen:2003kt}. \label{potential}}
\end{figure}
\end{center}

In Fig. \ref{potential}, one presents the line shapes of the
potentials listed in Eqs. (\ref{ss}) and (\ref{qq}). For
$Y^{s\bar{s}}$, the exchange potential of the $\phi$ meson can be
ignored compared with that of the $\eta$ meson. The total
effective potential of $Y^{s\bar{s}}$ is dominated by the $\eta$
exchange potential. For $Y^{u\bar{u}/d\bar{d}}$, the $\pi$ meson
plays an important role especially in the range of $r>5$
GeV$^{-1}$ since the exchange potentials of $\rho$, $\omega$,
$\sigma$ and $\eta$ decay exponentially with $r$. The behavior of
the potential depicted in Fig. \ref{potential} indicates that we
only need to consider the $\eta$ meson exchange potential for
$Y^{s\bar{s}}$ and the $\pi$ meson exchange potential for
$Y^{u\bar{u}/d\bar{d}}$ when finding the bound state solution by
solving Schr\"{o}dinger equation. Furthermore, whether there exist
bound state solutions for $Y^{s\bar{s}}$ and
$Y^{u\bar{u}/d\bar{d}}$ systems is closely related to the
corresponding strengths of the $D_{s0}(2317)D_s\eta$ and
$D_{0}(2400)D\pi$ couplings.

\begin{center}
\begin{figure}[htb]
\begin{tabular}{cc}
\scalebox{0.95}{\includegraphics{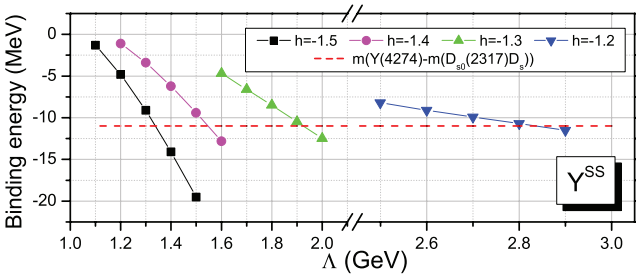}}
\end{tabular}
\caption{(Color online.) The obtained bound state solutions of
$Y^{s\bar{s}}$ system dependent on $h$ values and $\Lambda$. Here,
we also compare our result with the mass difference (red dashed
line) between $Y(4274)$ and the threshold of $D_{s0}(2317)D_s$.
\label{Ess}}
\end{figure}
\end{center}

In Fig. \ref{Ess}, we show the variation of the numerical result of
the bound state solutions for $Y^{s\bar{s}}$ with the values of
$h$ and $\Lambda$, which indicates that there indeed exists a
$D_{s0}(2317)\bar{D}_s+h.c.$ molecular charmonium corresponding to
newly observed enhancement $Y(4274)$. Our numerical results
overlap with the mass difference ($\sim-11$ MeV) between $Y(4274)$
and the threshold of $D_{s0}(2317)\bar{D}_s$. The corresponding
cutoff $\Lambda$ lies in a reasonable range which is expected to
be around 1-3 GeV. We also find that the larger $|h|$ values make
the corresponding $\Lambda$ become smaller, $i.e.$, $\Lambda$
tends to be around $1$ GeV, which is fully consistent with
the expected behavior of the potential of the S-wave
$D_{s0}(2317)\bar{D}_s+h.c.$ system.

Besides supporting the assignment of $Y(4274)$ as the S-wave
$D_{s0}(2317)\bar{D}_s+h.c.$ molecular state, our dynamical
calculation also provides a novel approach to extract the $h$
parameter, which encodes the important information of the
$D_{s0}(2317)D_s\eta$ interaction and the underlying properties of
$D_{s0}(2317)$ \cite{Aubert:2003fg}. This coupling can not be
extracted experimentally since the $D_{s0}(2317)\to D_s\eta$ decay
is forbidden kinematically. Our result indicates that the $|h|$
value corresponding to the binding energy of the S-wave
$D_{s0}(2317)\bar{D}_s+h.c.$ system consistent with mass
difference ($\sim-11$ MeV) is in the range $1.2\sim 1.5$
associated with reasonable $\Lambda$ value, which can be confirmed
by further theoretical study.

We extend the same formalism to the $Y^{u\bar{u}/d\bar{d}}$
system, where input parameter $h$ for the $D_0(2400)D\pi$ coupling
is constrained by the decay width of the $D_0(2400)\to D\pi$ to be
$h=-0.56\pm0.2$ \cite{Casalbuoni:1996pg}. The binding energy of
the $Y^{u\bar{u}/d\bar{d}}$ system is $-9.85$, $-10.11$, $-10.23$,
$-10.30$, $-10.34$, $-10.38$, $-10.42$ MeV corresponding to the
typical value of $\Lambda=0.9,\,1.0,\,1.1,\,1.2,\,1.3,\,1.4,\,1.5$
GeV, where the bound state solution of the $Y^{u\bar{u}/d\bar{d}}$
system is insensitive to $\Lambda$, which indicates the existence
of the molecular cousin of the S-wave $D_{s0}(2317)\bar{D}_s+h.c.$
molecular state, $i.e.$, an S-wave $D_0(2400)\bar{D}+h.c.$
molecular charmonium. Thus, finding the evidence of S-wave
$D_0(2400)\bar{D}+h.c.$ molecular charmonium can provide important
support to the assignment of $Y(4274)$ as an S-wave
$D_{s0}(2317)\bar{D}_s+h.c.$ molecular state. The important
hidden-charm decay mode of the S-wave $D_0(2400)\bar{D}+h.c.$
molecular charmonium is $J/\psi\omega$, which is the same as in
the case of $Y(3930)$ \cite{Abe:2004zs,Aubert:2007vj}.

\begin{center}
\begin{figure}[htb]
\begin{tabular}{cc}
\scalebox{0.75}{\includegraphics{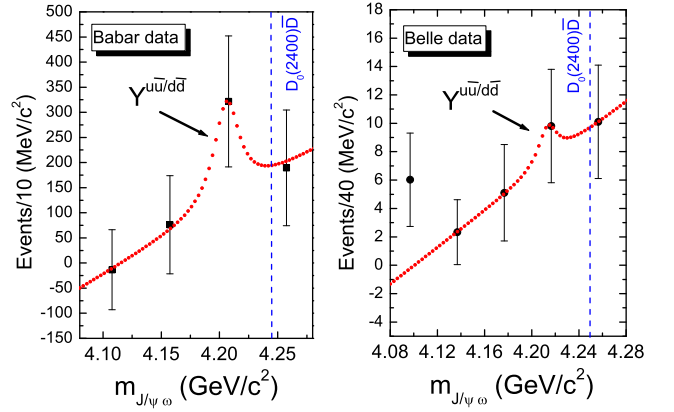}}
\end{tabular}
\caption{The $J/\psi\omega$ invariant mass spectrum in the $B\to
J/\psi\omega K$ decay announced by the Belle Collaboration
\cite{Abe:2004zs} and the Babar Collaboration
\cite{Aubert:2007vj}. Here, the vertical blue dashed line denotes
the threshold of $D_0(2400)\bar{D}$. The red dotted line is the
fitting result.  \label{4202}}
\end{figure}
\end{center}

From the published experimental data of the $J/\psi\omega$ invariant
mass spectrum \cite{Abe:2004zs,Aubert:2007vj}, we indeed notice an
enhancement structure around 4.2 GeV just below the threshold of
the $D_0(2400)\bar{D}$ pair as illustrated in Fig. \ref{4202},
which is amazingly consistent with our prediction of the S-wave
$D_0(2400)\bar{D}+h.c.$ molecular charmonium. We expect further
high-statistics measurement from future experiments to test our
prediction of the S-wave $D_0(2400)\bar{D}+h.c.$ molecular
charmonium.

The S-wave
$D_s0(2317)\bar{D}_s/D_0(2400)\bar{D}$ molecular state with
spin-parity $J^P=0^-$ does not couple to the $D\bar{D}
/D_s\bar{D}_s$ channels, which is strictly forbidden by the
conservation of the parity and angular momentum. In addition, the
S-wave $D_s0(2317)\bar{D}_s/D_0(2400)\bar{D}$ molecular state may
couple to the $D^*\bar{D}/D_s^*\bar{D}_s$ and
$D^*\bar{D}^*/D_s^*\bar{D}_s^*$ via P-wave, which is expected to
be suppressed compared to the S-wave mode. Due to the above
reasons, the coupled-channel effect on the S-wave
$D_s0(2317)\bar{D}_s/D_0(2400)\bar{D}$ molecular state may be
weak, which is ignored in this work.

As an S-wave $D_{s0}(2317)\bar{D}_s+h.c.$ molecular charmonium
with $J^P=0^-$, the decay modes of $Y(4274)$ include the
hidden-charm decay mode $J/\psi\phi$ observed by CDF
\cite{collaboration:2010aa}, the two-body P-wave open-charm decays
$D_s\bar{D}_s^*+h.c.$ and $D_s^*\bar{D}_s^*$, the radiative decay
$D_s^*\bar{D}_s\gamma+h.c.$, and the iso-spin violating three-body
strong decay $D_{s}\bar{D}_s\pi^0$ via the $\eta-\pi^0$ mixing
mechanism \cite{Aaltonen:2009tz,Abe:2004zs}. Similarly
$Y^{u\bar{u}/d\bar{d}}$ can decay into $J/\psi\omega$,
$D\bar{D}^*+h.c.$, $D^*\bar{D}^*$, $D\bar{D}\pi$,
$D^*\bar{D}\gamma+h.c.$ etc.

After figuring out the underlying structure of $Y(4274)$ and predicting its molecular cousin, we notice that there exist two event clusters around the ranges of $\Delta M\sim 1.27$
GeV and $1.4<\Delta M<1.5$ GeV marked by yellow and pink
in Fig. \ref{BW}, if we focus on the remaining CDF's data corresponding to
$\Delta M>1.24$ GeV. If these two event clusters are confirmed
by future experiments, we
might also try to understand them under the same framework of the
molecular charmonium. Basing on the present low-statistic data
\cite{collaboration:2010aa}, we speculate that the structure
appearing at $\Delta M\sim 1.27$ is related to the
$D_s\bar{D}_{s1}^\prime(2460)$ or $D_s^*\bar{D}_{s0}(2317)$
system. The other one in the range $1.4<\Delta M<1.5$ GeV may
result from the $D_{s}\bar{D}_{s1}(2536)$,
$D_{s}\bar{D}_{s2}(2573)$, $D_{s}^*\bar{D}_{s1}^\prime(2460)$ and
$D_{s}^*\bar{D}_{s1}(2536)$ systems since the event cluster in the
range $1.4<\Delta M<1.5$ GeV just overlaps with the corresponding
thresholds (see Fig. \ref{BW} for more details). One may recall
the similar situation before finding the evidence of $Y(4274)$ by
CDF \cite{collaboration:2010aa}. The CDF's data with an integrated
luminosity of 2.7 fb$^{-1}$ reported in Ref.
\cite{Aaltonen:2009tz} only displayed the event cluster at 4.27
GeV besides the evidence of $Y(4140)$. Confirming the above
speculation by further experimental study of $J/\psi\phi$
invariant mass spectrum from $B$ decay will not only test the
molecular charmonium assignments of $Y(4140)$ and $Y(4274)$, but
also improve our understanding of the line shapes appearing at
hidden-charm invariant mass spectra.

In summary, the newly observed structure $Y(4274)$ in the
$J/\psi\phi$ invariant mass spectrum is first interpreted as the
S-wave $D_s\bar{D}_{s0}(2317)+h.c.$ molecular charmonium well from the
dynamical study of the system composed of the pseudoscalar and
scalar charmed mesons. Furthermore, we predict the S-wave $D\bar{D}_{0}(2400)+h.c.$
molecular charmonium  appearing as the cousin of $Y(4274)$, which is
consistent with the enhancement structure around 4.2 GeV in the
$J/\psi\omega$ invariant mass spectrum from $B$ decay
\cite{Aaltonen:2009tz,Abe:2004zs}. Thus, the enhancement
structures including the present $Y(4274)$, the previous $Y(4140)$
and $Y(3930)$ observed in the $J/\psi\phi$
\cite{collaboration:2010aa,Aaltonen:2009tz} and $J/\psi\omega$
\cite{Aaltonen:2009tz,Abe:2004zs} invariant mass spectra respectively, can be
accommodated well in the uniform framework of the molecular
charmonium. In addition, we find two possible event clusters in the
$J/\psi\phi$ invariant mass spectrum might related to the
molecular charmonia, which can be tested by high-statistic
experimental data in future experiment.

\section*{Acknowledgment}

\noindent X.L. thanks Dr. Jun He for his help of fitting data.
This project is supported by the National Natural Science
Foundation of China under Grants No. 10705001, No. 11035006, No. 11047606, No.
10947204, No. 10625521, No. 10721063, the Ministry of Science and
Technology of China (2009CB825200), and the Ministry of Education
of China (FANEDD under Grant No. 200924, DPFIHE under Grant No.
20090211120029, NCET under Grant No. NCET-10-0442, the Fundamental
Research Funds for the Central Universities under Grant No.
lzujbky-2010-69).

\vfil

\end{document}